\documentclass[slac_one]{revtex4}
\usepackage{graphicx}
\usepackage{fancyhdr}
\begin{document}

\title{{\small{2005 International Linear Collider Workshop - Stanford,
U.S.A.}}\\ %% Please keep this conference title here
\vspace{12pt}
Spatial Resolution Studies with a GEM-TPC in High Magnetic Fields} %% Paper title goes here


\author{Peter Wienemann}
\affiliation{DESY, Notkestr.~85, D-22607 Hamburg, Germany}
%
%\author{P. Lucas}
%\affiliation{FNAL, Batavia, IL 60510, USA}

\begin{abstract}
A large volume Time Projection Chamber (TPC) has been proposed
as main tracking device at the International Linear Collider (ILC).
Gas electron multipliers (GEMs) are studied as potential replacement
of the conventional wire based gas amplification system of TPCs.
This talk presents recent results from R\&D activities with a small
GEM-TPC prototype. The spatial resolution was measured for different
magnetic fields up to 4 T.
\end{abstract}

\maketitle

\thispagestyle{fancy}


\section{INTRODUCTION} % Section title should be in all capitals.

The ambitious physics program at the International Linear Collider
(ILC) poses stringent requirements on the performance of the detector.
An accurate momentum measurement and a good particle identification
relies crucially on precise tracking information.  Therefore the
development of the tracker needs special attention.  As main tracker
for a detector at the ILC, a large Time Projection Chamber (TPC) is
being studied. It allows the instrumentation of a large volume with
many voxels and represents only a minimum amount of material before
the calorimeters. Moreover it has good particle identification
capabilities, a genuine three-dimensional track reconstruction without
ambiguities, and concentrates its sensitive parts to the endplates,
allowing easy maintainability.  Contrary to previous TPCs with
multi-wire proportional chambers (MWPCs) for gas amplification, future
TPCs are likely to make use of Micro Pattern Gas Detectors
(MPGDs). One promising MPGD candidate is the Gas Electron Multiplier
(GEM)~\cite{ref:GEM}. Among its advantages are amplification
structures of order 100~$\mu$m giving rise to only tiny $\vec{E}
\times \vec{B}$ effects, a fast and narrow electron signal and
intrinsic ion backdrift suppression.

\section{THE DESY TPC PROTOTYPE}

In order to investigate the potential of TPCs with GEM amplification,
a small prototype has been built at DESY. The chamber has a maximal
drift length of 800~mm and a diameter of 270~mm. Its size has been
chosen such that it fits into a superconducting 5 T magnet available
for detector R\&D studies at DESY. The chamber endplate is equipped
with 24 $\times$ 8 = 192 readout pads of size 2.2 $\times$ 6.2 mm$^2$.
Two different pad layouts are investigated: First a layout
where the pads in each row are shifted by half a pitch with respect to
the pads in the two neighboring rows (staggered) and a second setup
with aligned pads (non-staggered).  The maximal drift length amounts
to 670~mm. Gas amplification is provided by a triple GEM structure
with two 2~mm transfer gaps and a 3~mm induction gap.  The readout
electronics is based on modules developed for the ALEPH experiment at
LEP.

\section{MEASUREMENTS IN HIGH MAGNETIC FIELDS}

One of the most important quantities of a TPC is the achievable
spatial resolution. It depends on various chamber parameters such as
the diffusion of the chosen gas, the pad size, the electronics, the
gas amplification settings, etc. Since the transverse diffusion
coefficient of gases strongly depends on the magnetic field, it is
necessary to perform spatial resolution measurements in magnetic
fields in order to get reliable estimates of the performance of the
final detector. A good quantity to compare spatial resolutions of
different small prototypes and to extrapolate to a large-scale device
is the single point resolution.

To find out what single point resolution might be feasible, a series
of measurements was carried out with cosmic muons at various magnetic
fields up to 4 T, the value proposed in the technical design report
for TESLA~\cite{ref:TESLA}. These runs were performed for two
different gases, namely Ar-CH$_4$-CO$_2$ (93-5-2) and Ar-CH$_4$ (95-5).

\subsection{Analysis Technique}
\label{sec:technique}

The reconstruction of tracks is done in three steps. First
three-dimensional space points are reconstructed from the pulses in
each row. In a second step, these space points are combined to tracks
using a three-dimensional track following algorithm.  Finally the
track parameters are fitted using a maximum likelihood fit which takes
the pad response function into account~\cite{ref:fit}.  To determine
the spatial resolution, the following procedure is applied: One row is
chosen and the horizontal track position is re-fitted using only data
from that row keeping all other track parameters (inclination, width
and curvature) fixed to the values obtained from a fit to all pad
rows. The distribution of the difference of this re-fitted horizontal
position and the original horizontal position is stored. Subsequently
the same method is repeated with the only modification that the fixed
parameters are set to the values obtained from a track fit excluding
the information from the chosen row. This is done for all rows. A good
estimate of the spatial resolution is obtained by calculating the
geometric mean of the widths of the two distributions determined in
the described way.

\subsection{Diffusion Coefficient}

\begin{figure*}[t]
\centering
\includegraphics[width=80mm]{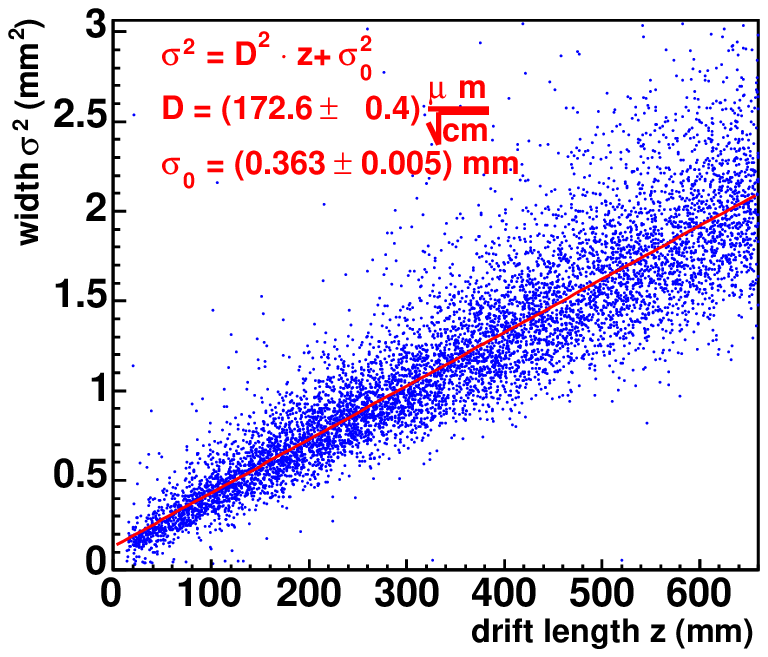}
\includegraphics[width=80mm]{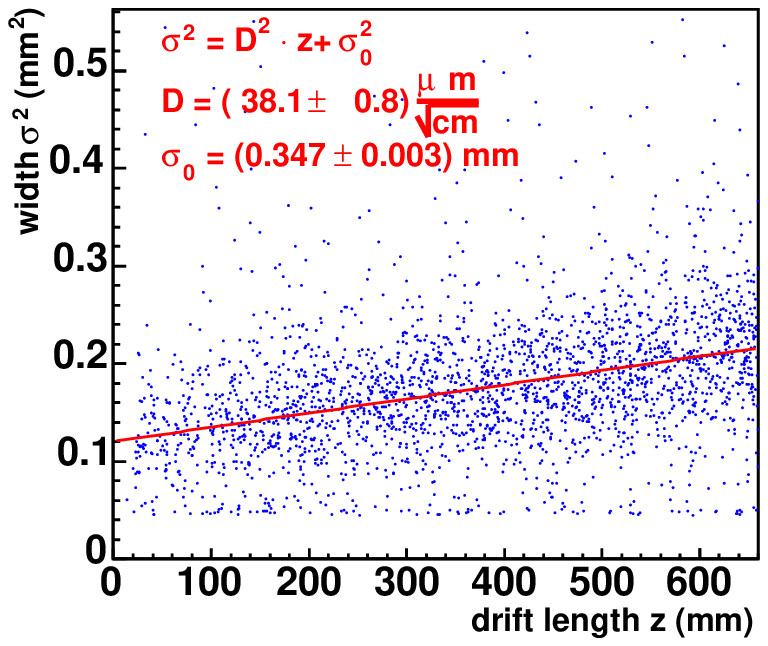}
\caption{Square of charge cloud width on the pads versus drift length
         for Ar-CH$_4$-CO$_2$ (93-5-2) at 1~T (left) and for Ar-CH$_4$ (95-5) at 4~T (right).}
\label{fig:widthVsDrift}
\end{figure*}

As mentioned above, diffusion is an important factor influencing the
spatial resolution of a TPC. It leads to a spread of the primary
charge cloud due to collisions with gas atoms/molecules. Therefore the
width $\sigma$ of the charge cloud on the pads increases with
increasing drift distance $z$:
\begin{equation}
    \sigma^2 = D^2 z + \sigma_0^2 .
\label{eq:width}
\end{equation}
$D$ is the diffusion constant and $\sigma_0$ the defocussing term describing
the charge widening in the amplification system. $D$ varies with the
magnetic field according to the formula
\begin{equation}
    \frac{D(B)}{D(0)} = \frac{1}{\sqrt{1 + \omega^2 \tau^2}}
\label{eq:diffusion}
\end{equation}
where $\omega = eB/m$ is the cyclotron frequency and $\tau$ the mean time
between collisions.

\begin{figure*}[t]
\centering
\includegraphics[width=120mm]{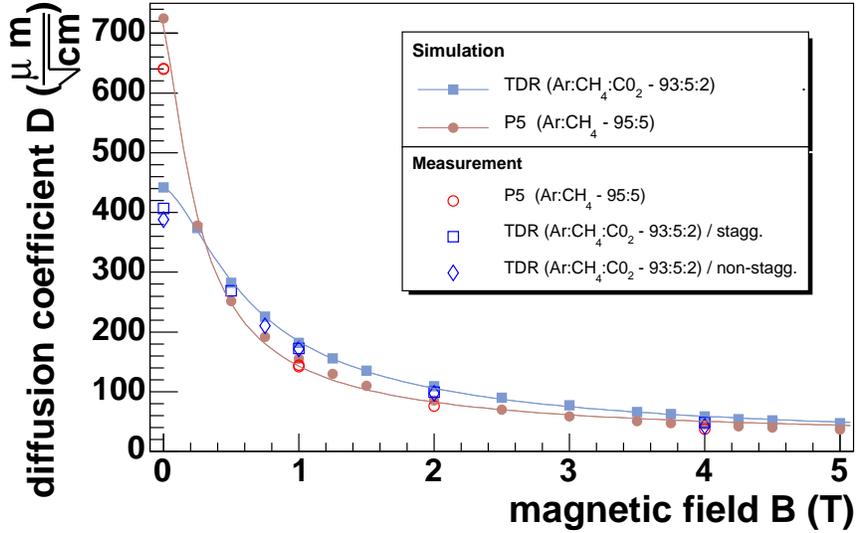}
\caption{Diffusion coefficient versus magentic field for Ar-CH$_4$-CO$_2$ (93-5-2) and for Ar-CH$_4$ (95-5).
         Both the measurements and the values from a Garfield simulation are shown.}
\label{fig:diffusion}
\end{figure*}

Figure~\ref{fig:widthVsDrift} shows the square of the fitted charge
cloud width versus the drift length for two examples, Ar-CH$_4$-CO$_2$
(93-5-2) at 1~T and for Ar-CH$_4$ (95-5) at 4~T. The data is well described by
Equation~\ref{eq:width}. The diffusion coefficient and the defocussing
terms are obtained from a linear fit to the data. The results are
shown in the plots. Such measurements are accomplished for various
magnetic fields for both gases. The outcome is shown in
Figure~\ref{fig:diffusion} where the diffusion coefficient is plotted
versus the magnetic field. In addition to the measurements, results
from a Garfield simulation \cite{ref:Garfield} are included. The
diffusion coefficient drops in accordance with
Equation~\ref{eq:diffusion} with the magnetic field. Qualitative
agreement is achieved between measurement and simulation, although
quantitatively the simulation seems to provide systematically slightly
higher values than the measurement. This phenomenon has been observed
by various groups~\cite{ref:discrepancy}.

\subsection{Spatial Resolution}

\begin{figure*}[t]
\centering
\includegraphics[width=100mm]{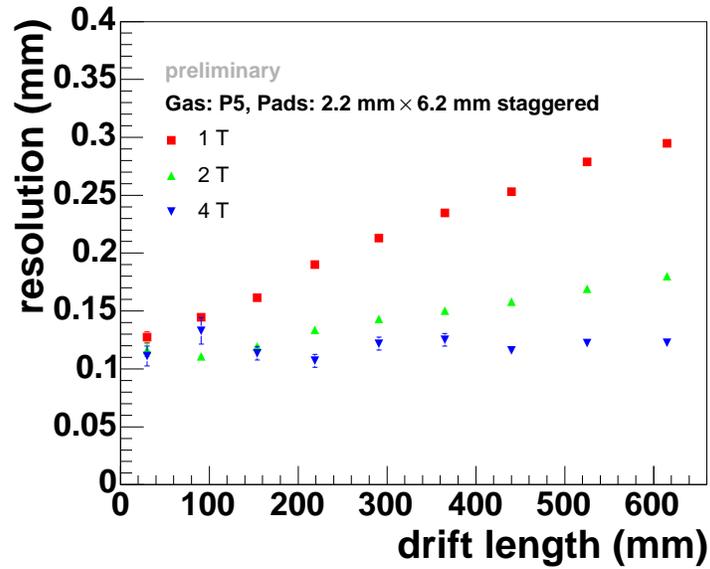}
\caption{The transverse resolution versus drift length for Ar-CH$_4$ (95-5) gas.}
\label{fig:resolution}
\end{figure*}

Using the procedure described in Section~\ref{sec:technique}, the
transverse single point resolution is determined for various magnetic
fields.  Figure~\ref{fig:resolution} shows the results for Ar-CH$_4$
(95-5) as a function of the drift length for 1~T, 2~T and 4~T. Due to
diffusion the spatial resolution gets worse for increasing drift
length. This effect is significantly suppressed for high magnetic
fields because of the reduced diffusion coefficient. For 2.2 $\times$
6.2 mm$^2$ pads, the current preliminary analysis yields about 120
$\mu$m transverse resolution for Ar-CH$_4$ (95-5) at 4~T.  This is in
full agreement with the requirements listed in \cite{ref:TESLA}.
Nevertheless further studies are under way to gain a deeper
understanding of the systematics involved in the reconstruction
method.

\section{CONCLUSIONS}

A small TPC prototype with GEM foils for gas amplification has been
successfully built to measure the spatial resolution in high magnetic
fields. Cosmic muon runs were carried out in B fields up to 4 T. For
2.2 $\times$ 6.2 mm$^2$ pads, the present preliminary analysis yields
about 120 $\mu$m transverse resolution with Ar-CH$_4$ (95-5) at 4~T, in full
agreement with the TESLA TDR requirements. These are encouraging
results revealing the potential of GEMs as TPC gas amplification
system.


\begin{thebibliography}{9}   % Use for  1-9  references
%\begin{thebibliography}{99} % Use for 10-99 references

\bibitem{ref:GEM}
  F.~Sauli, ``GEM: A new concept for electron
  amplification in gas detectors'', {\it Nucl.~Instrum.~Methods},
  A 386, 1997.

\bibitem{ref:TESLA}
  T.~Behnke, S.~Bertolucci, R.-D.~Heuer, and
  R.~Settles, ``TESLA Technical Design Report'', DESY, Hamburg, Germany,
  DESY 2001-011 and ECFA 2001-209, 2001.

\bibitem{ref:fit}
  D.~Karlen, ``Pad geometry study for a linear collider TPC'', published in
  the proceedings of the International Workshop on Linear Colliders (LCWS 2002),
  Jeju Island, Korea, 2002.

\bibitem{ref:Garfield}
  R.~Veenhof,``Garfield - simulation of gaseous detectors'',
  {\tt http://cern.ch/garfield}.

\bibitem{ref:discrepancy}
  See e.~g.~A.~M\"unnich, ``R\&D for a TPC with GEM Readout'', talk given at
  International Linear Collider Workshop 2005, Stanford, USA and
  D.~Karlen, ``GEM-TPC performance in magnetic field'', talk given at
  International Linear Collider Workshop 2004, Paris, France.

\end{thebibliography}
\end{document}